# Topological charge density around static color sources[*][†]

Manfried Faber, Harald Markum, Štefan Olejník[‡] and Wolfgang Sakuler

Institut für Kernphysik, Technische Universität Wien, A–1040 Vienna, Austria

We analyze the topological structure of quenched $QCD$ in the presence of static color sources. Distributions of the topological charge density around static quarks and mesons are computed in both phases of $QCD$. We observe a suppression of topological fluctuations in the vicinity of external sources. In the confinement phase, the suppression occurs in the whole flux tube between the static quark and antiquark.

**1.** The physical mechanism of quark confinement in $QCD$ is still not completely understood. It is generally believed, however, that it is due to non-trivial topological properties of the theory.

To get some insight into the confinement mechanism and the corresponding vacuum structure it is instructive to study various physical quantities locally around color sources. In previous works we investigated the chiral condensate, charge density, color field distribution, etc., around static sources in full $QCD$ at finite temperature [1]. We observed that the vacuum structure changes significantly around external charges. Our results agree with phenomenological models like the bag model where the vacuum inside the bag is of perturbative nature and differs from the nonperturbative medium outside.

In the present work we focus on the topological properties of $QCD$. We study topological fluctuations around static quarks and static mesons in both phases of quenched $QCD$ by computing correlation functions between Polyakov loops and the topological charge density.

**2.** There is no unique way of defining a lattice operator that converges in the naive continuum limit to the continuum expression for topological charge. We have chosen two possibilities, the "plaquette" ($P$) and the "hypercube" ($H$) definition [2], in which the topological charges $Q^{(P,H)}$ are defined as integrals (sums) of topological charge densities:

$$q^{(P,H)}(x) = -\frac{1}{2^4 32\pi^2} \sum_{\mu,\ldots=\pm 1}^{\pm 4} \tilde{\epsilon}_{\mu\nu\rho\sigma} \operatorname{Tr} \mathcal{O}_{\mu\nu\rho\sigma}^{(P,H)}, \quad (1)$$

where operators $\mathcal{O}_{\mu\nu\rho\sigma}^{(P,H)}$ are (in usual notation)

$$\mathcal{O}_{\mu\nu\rho\sigma}^{(P)} = U_{\mu\nu}(x) U_{\rho\sigma}(x), \quad (2)$$

$$\begin{aligned}\mathcal{O}_{\mu\nu\rho\sigma}^{(H)} &= U_\mu(x) U_\nu(x+\hat{\mu}) U_\rho(x+\hat{\mu}+\hat{\nu}) \\ &\times U_\sigma(x+\hat{\mu}+\hat{\nu}+\hat{\rho}) U_\mu^\dagger(x+\hat{\nu}+\hat{\rho}+\hat{\sigma}) \\ &\times U_\nu^\dagger(x+\hat{\rho}+\hat{\sigma}) U_\rho^\dagger(x+\hat{\sigma}) U_\sigma^\dagger(x). \quad (3)\end{aligned}$$

At finite coupling $\beta$ and finite lattice spacing neither of the above definitions yields inte-ger topological charges. This is known to be due to renormalization factors relating the continuum and lattice operators of topological charge [3]. This problem can be overcome by using a cooling procedure. Cooling is subsequent minimizing of the action with respect to all individual links. We employ a slightly modified cooling method from Ref. [4]. Within a cooling step, we minimize in all three diagonal $SU(2)$ subgroups of $SU(3)$, in random order, and finally check after any minimization step that the changes of links are small enough not to destroy immediately the topological charge. After a few cooling steps, topological charges of lattice configurations, determined using both definitions, agree and are close to integer values.

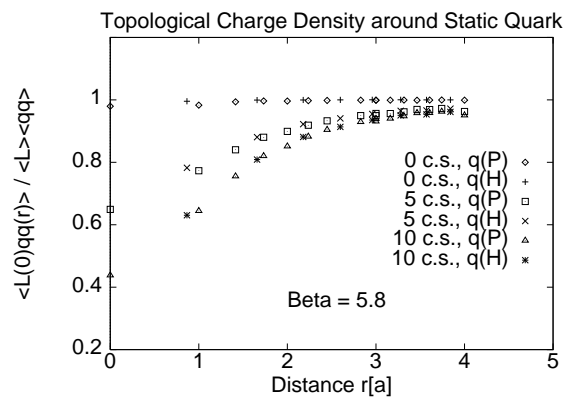

Figure 1. Squared topological charge density around a static quark after 0, 5, and 10 cooling steps at $\beta = 5.8$ (deconfinement phase).

[*]Presented by W. Sakuler.
[†]Supported in part by FWF (Contract No. P9428–PHY), and by BMWF.
[‡]On leave from Institute of Physics, Slovak Academy of Sciences, SK–842 28 Bratislava, Slovakia.

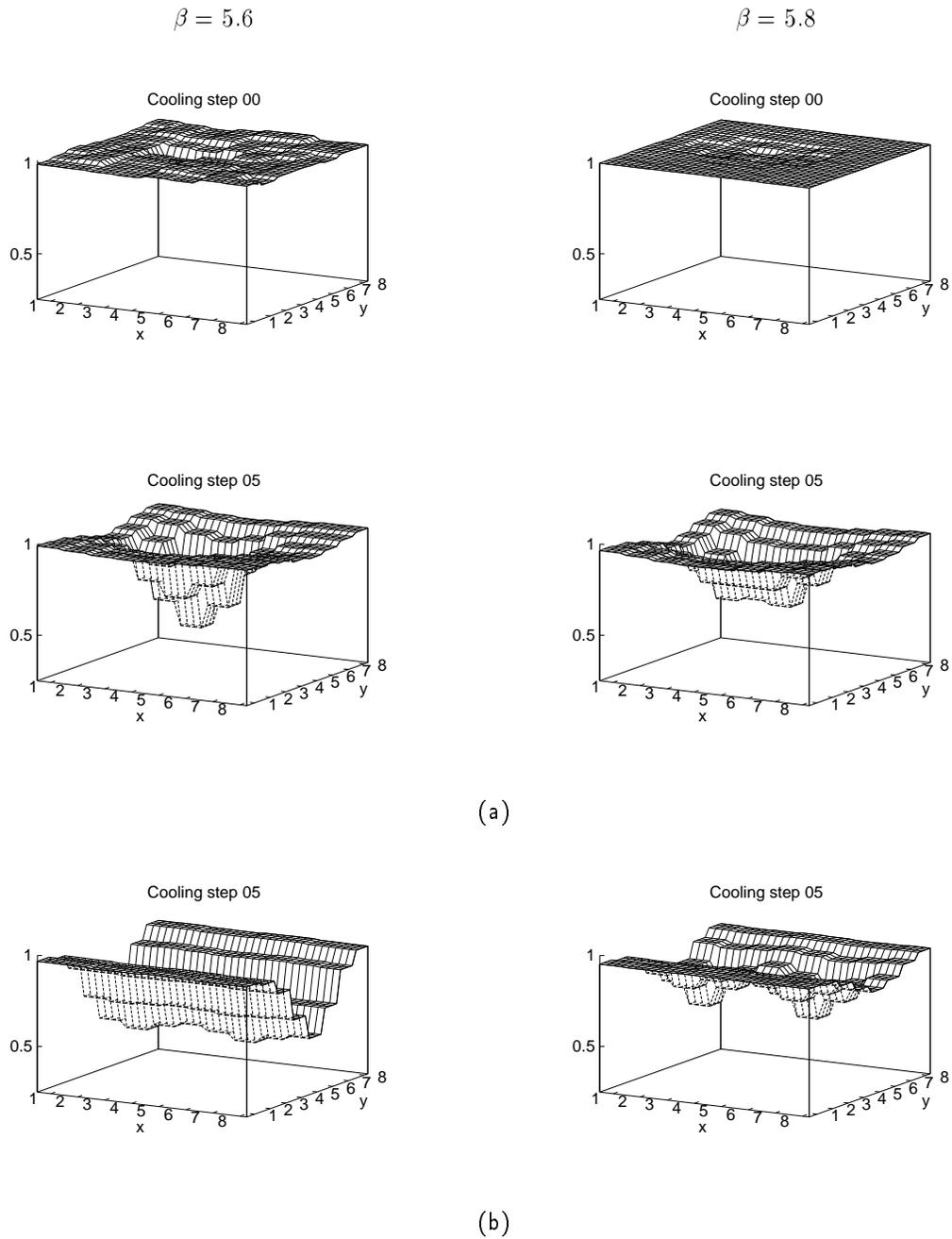

Figure 2. Squared topological charge density around a static quark-antiquark pair at $\beta = 5.6$ and $\beta = 5.8$: (a) $d = 2$, after 0 and 5 cooling steps, (b) $d = 4$, after 5 cooling steps.

A static quark is represented by the Polyakov loop $L(r)$ which describes the propagation of a charge with infinite mass. To obtain the local distribution of the topological charge density around single quarks and around a meson with quark-antiquark separation $d$ we calculate the correlation functions $\langle L(0) q^2(r) \rangle$ and $\langle L(0) L^\dagger(d) q^2(r) \rangle$, respectively, and normalize them to the corresponding vacuum expectation values. We take the square of the topological charge density as the simplest choice leading to nontrivial correlations.

**3.** The simulations were performed on an $8^3 \times 4$ lattice with periodic boundary conditions using a Metropolis algorithm. We evaluated path integrals in pure $SU(3)$ with standard Wilson action both in the confinement ($\beta = 5.6$) and in the deconfinement phase ($\beta = 5.8$). We made $10^5$ iterations per $\beta$ and measured our observables after every 50th iteration. Each of these 2000 configurations was subjected to 50 cooling steps.

In Fig. 1 the square of the topological charge density around a single static quark in the deconfinement phase is shown after 0, 5, and 10 cooling steps, for both definitions of the density. Without cooling practically no effect is visible, but cooling soon uncovers a suppression of the density. Results from both definitions of the density agree after a few cooling steps. In the confinement phase one expects $\langle L(0) q^2(r) \rangle = 0$, what was confirmed by our measurements. This reflects the fact that a single quark is a forbidden state in the confinement phase of pure gauge $QCD$.

In Fig. 2 distributions of the topological charge density (squared) around a meson are presented for both phases of pure $SU(3)$. The data for 0 and 5 cooling steps are visualized at $q\bar{q}$-distance $d = 2$ in Fig. 2a. For 5 cooling steps the suppression becomes again clearly visible. The results indicate a stronger effect in the confinement regime. The difference between both phases is more striking in Fig. 2b where plots for the 5th cooling step and larger separation, $d = 4$, are drawn. The valley of suppressed topological charge density in the confinement phase reflects the formation of a flux tube between the static quark-antiquark pair. In the deconfinement phase, no flux tube is visible and the effect is peaked near the position of the sources (cf. Fig. 1).

**4.** Our results show a change of the topological properties of lattice vacuum configurations across the deconfining phase transition. Configurations with nonvanishing topological charge appear very seldom in the deconfinement phase. The topological charge density is lowered in the vicinity of external color sources in both phases of quenched $QCD$. However, the effect is more pronounced in the confinement phase where it is observed in the whole flux tube between the static quark and antiquark. A similar investigation for $QCD$ with dynamical fermions is in progress.